\documentstyle[11pt]{article}
\input psfig
\long\def\comment#1{}
\begin{document}
\title{On Rotating a Qubit}
\author{Subhash Kak\\
Department of Electrical \& Computer Engineering\\
Louisiana State University\\
Baton Rouge, LA 70803-5901; {\tt kak@ee.lsu.edu}}
\maketitle

\begin{abstract}
The state function of a quantum object is undetermined with
respect to its phase. This indeterminacy does not matter
if it is global, but what if the components of the state
have unknown relative phases? Can useful computations be performed
in spite of this local indeterminacy? We consider this question
in relation to the problem of the rotation of a qubit
and examine its broader implications for quantum computing.

\end{abstract}


\subsection*{Introduction}

Quantum mechanics provides us a means of extracting information
about a physical system, but this information depends 
on the manner in which the system is prepared
and the measurement apparatus deployed.
The state function of the system is defined on
the complex plane whereas the observations can
only be real, which means that the state function
cannot be completely known.

Quantum computing algorithms as visualized now proceed
with the register in a pure state.
Normally, this state is taken to be the
all-zero state
of $n$-qubits: $| 0\rangle| 0\rangle ...| 0\rangle,$
or $2^n$ amplitudes $(1, 0, 0... 0)$, which
by a process of rotation transformations 
on each qubit is transformed into the 
state with amplitudes $(\frac{1}{\sqrt N},\frac{1}{\sqrt N},\frac{1}{\sqrt N}
... \frac{1}{\sqrt N}),$ where $N = 2^n $.
It is implicitly assumed that the phase uncertainty in each
of these states is global, and so it can be ignored.

Is this assumption realistic?
Can the cost and realizability of the
initial pure state and its rotations to the 
superposition state be estimated?
This problem was considered in
a recent paper \cite{Ka99}, where the author discussed its
implications 
for initializing the state of a quantum
register.
We return to this problem here to clarify several
points related to preparation of pure states and to focus on the
difficulty of controlling the evolution of a quantum state,
which is essential to do in a useful quantum computation.

\subsection*{The cost of preparing a pure state}

A pure state is one which yields a specific outcome in 
a given test designed to elicit the maximum number
of outcomes associated with the system \cite{Pe95}.
Examples are Stern-Gerlach experiment for spin or
the use of a calcite crystal for photon polarization.
It is reasonable to represent the all-zero state by
$(1, 0, 0...0),$ if it is assumed that each of the
qubits has been prepared identically and there is
no dynamical evolution. In other words, it is assumed
that qubits emerging out of the state preparation
apparatus are frozen in their state and installed
at the appropriate locations in the quantum register.

It is important to note that a dynamically
evolving pure state remains a pure state, since
pure states are analogous to unit vectors along
a set of orthogonal basis vectors, and any
vector in the N-dimensional space can be taken
to be one of the basis vectors. 

From an information-theoretic point of view, 
a transition from the initial $(1, 0, 0...0)$ state to the
$(1, 1, 1...1)$ superposition state is equivalent to a transition
to some other $(z_1 , z_2 ...z_N )$ state (where $z_i$s are arbitrary
complex numbers and it is assumed that the amplitudes are
to be normalized), because each of the final states
will be obtained by the application of a suitable 
unitary transformation.
Many authors have assumed that
the latter state, with the unknown $z$ weights, is {\it more} ``complex''
in some fundamental way.
The reason behind this mistaken view is the fact that
it is easy to obtain the superposition state
$(1, 1, 1...1)$
mathematically by the use of elegant rotation transformations on 
the qubits, individually \cite{Sh97,Gr97}?

The $n$-qubit vector is a tensor product of the $n$ individual
qubit vectors. As there is no reason to assume that the
relative phase between the superposed states is zero,
the individual qubits should be written as 
$ z_{i1} | 0\rangle + z_{12} | 1\rangle .$
The $n$-qubit state will then have $N-1$ unknown relative
phases and one global unknown phase.

It is common to ignore the relative phases in state
vectors for ensembles, because in repeated measurements
the unknown phases can be assumed to average out to zero.
However, in quantum computing we speak about individual
quantum system and so this averaging out is not
permissible.

In entangled states the phases do have definite relationships.
But the qubits that go into the formation of the 
$n$-qubit state vector are not permitted to be entangled
for computationally useful operations to be performed on them.

\subsection*{Preparing a qubit}

For simplicity, we consider a qubit to have
the form $| \phi\rangle = \alpha e^{i \theta_1} | 0\rangle + \beta  e^{i \theta_2}| 1\rangle,$
where $ \alpha, \beta \in R$ and $\alpha^2 + \beta^2 = 1$.
A qubit may be prepared by starting with an available 
component or
superposition state and 
transforming it into the desired superposition state by
applying a unitary transformation.

The simplest way to prepare a pure state is to subject qubits
to a test and discard all the qubits that do not yield the
desired outcome. Pure states are unit vectors along a set
of orthogonal axes, and two examples are $ e^{i \theta_1} | 0\rangle$ or
$ e^{i \theta_2} | 1\rangle$ at angles of 0 or 90 degrees; other
orthogonal axes can likewise be chosen. The standard basis
observables are $ | 0\rangle$ and $ | 1\rangle$.

To consider different ways of initializing 
qubits,
assume that the starting pure state is $ | 0\rangle$, 
determined excepting an arbitrary phase angle. One can obtain
a superposition state by use of a rotation operator which, in its
most general form, is the following matrix

\begin{equation}
			 \left[ \begin{array}{cc}
                                  \alpha e^{i \theta_1}  & \beta e^{-i \theta_2} \\
                                  \beta e^{i \theta_2} & -\alpha e^{-i \theta_1} \\
                               \end{array} \right] 
\end{equation}
where $\theta_1$ and $\theta_2$ are unknown phase angles. This leads to the 
superposition state

\begin{equation}
 \alpha e^{i \theta_1}  | 0\rangle + \beta e^{i \theta_2} | 1\rangle 
\end{equation}

Here we ignore the fact that
the initial state will have some uncertainty associated 
with it due to 
the impossibility of a perfect implementation of the
angle of the basis vector. Furthermore, the presence of
noise and entanglement with the
environment imply that the superposition state
will not be completely controllable, and these questions
are also ignored here. If it can be arranged that
$ \theta_1 = \theta_2 ,$ then any appropriate
rotation can be achieved.
But we have no means of ensuring
this equality, excepting by the use
of entanglement which would defeat our purpose.

If we rely on the
simpler method of starting with a pure state as a
unit vector in the 45 degree direction, such a pure state,
when viewed from the axes at 0 and 90 degrees, will
be defined  as

\begin{equation}
\frac{1}{\sqrt 2} (e^{i \theta_1}  | 0\rangle + e^{i \theta_2} | 1\rangle )
\end{equation}

This pure state will resolve into the basis states in the
directions of 0 and 90 degrees with equal probability.

Is there a way to rotate a qubit by any specific
angle?
For convenience assume that the operator

\begin{equation}
M = \frac{1}{\sqrt 2} \left[ \begin{array}{cc}
                                  1 & 1 \\
                                  1 & -1 \\
                               \end{array} \right] 
\end{equation}
is implementable. 
When applied to the qubit $  \frac{1}{\sqrt 2}( |0\rangle +  |1\rangle),$
it will lead to the pure state $|0\rangle$.
But since the qubit should be realistically seen to be
$\frac{1}{\sqrt 2} (e^{i \theta_1}  | 0\rangle + e^{i \theta_2} | 1\rangle ),$
an operation by M will take the qubit only to

\begin{equation}
\frac{(e^{i \theta_1} + e^{i \theta_2})}{2} | 0\rangle  +
\frac{(e^{i \theta_1} - e^{i \theta_2})}{2} | 1\rangle
\end{equation}

The probability of obtaining a $|0\rangle$ will now
be 
$ \frac{1}{2} [1 + cos ( \theta_1 - \theta_2)],$
whereas the probability of obtaining a $ | 1\rangle$ will be
$ \frac{1}{2} [1 - cos ( \theta_1 - \theta_2)].$
The probabilities for the basis observables are
not exactly $\frac{1}{2}$, and they depend on the
starting {\em unknown} $\theta$ values.
Thus, the qubit can end up anywhere on the unit circle.
As example, consider $\theta_2 = 0, ~\theta_1 = \pi /2$,
the probabilities of $ | 0\rangle $ and $ | 1\rangle$
will remain $\frac{1}{2}$ even after the 
unitary transformation has been applied!

If one assumes that $ \theta_1 - \theta_2 = \theta_d$
is uniformaly distributed over 0 to $\pi,$ the probability
of obtaining $| 0\rangle$ turns out to be $\frac{1}{ \pi \sqrt { y - y^2 }},$
where $y$ ranges from 0 to 1.
The expected value of this continues to be $\frac{1}{2}.$

If we consider the rotation of qubit (2) by the unitary operator (4),
we get the following probabilities
$ \frac{1}{2} [1 + \alpha \beta cos ( \theta_1 - \theta_2)],$
$ \frac{1}{2} [1 - \alpha \beta cos ( \theta_1 - \theta_2)],$
for obtaining the states $| 0\rangle$ and $ | 1 \rangle$
respectively.
In other words, the fundamental phase uncertainty makes
it impossible to calibrate rotation operators.
The operator (4) is unable to rotate the pure state (3)
by 45 degrees.

Since the rotation of qubits is visualized to be done in stages in
the currently conceived implementation schemes, any of
these gates can introduce phase uncertainty that 
will be impossible to compensate for in the ongoing
unitary operations.

\subsection*{Conclusions}

Rotation operations are basic for the implementation of the currently
envisaged quantum algorithms. Lacking information
regarding phase of the qubit, it
is clear that these operators will not work correctly.
Furthermore,
each gate will introduce its own random phase
uncertainty because the operation $A | \phi\rangle$
and $A e^{i \theta} | \phi\rangle$ are indistinguishable.
As example of this, consider the unitary
operator
$
 \left[ \begin{array}{cc}
                                  1 & 0 \\
                                  0 & -1 \\
                               \end{array} \right] 
$
which has been proposed to make the relative phases
of $|0\rangle$ and $|1\rangle$ flip. Even assuming
that the original qubit was $ \frac{e^{i\theta}}{\sqrt 2} (|0\rangle + |1\rangle),$ the new phases become $\theta$ and $\theta + \pi,$ which
cannot be characterized phase reversal. Since the initial
phases have uncertainty, one cannot steer the qubit to a
desired change.

Error correcting codes cannot be used to correct this
uncertainty because it is without bound.
Furthermore, if error correction worked for this
problem that would allow for us to make the
relative phases zero, which gets us more than
what quantum theory allows. 

On the other hand, if the initial phase uncertainty can be
lumped together then an appropriate sequence of unitary
transformations will, in principle, steer the state to the
desired solution. 
The challenge then will be the physical
implementation of this transformation in a manner so that
this lumped uncertainty doesn't diffuse to the
constituent qubits in an uncontrolled way.
But even if this could be done,
does the lumping together of the phase uncertainty 
imply limitations with respect to implementation
that will drastically reduce the advantages of
quantum computers?

\comment{
\subsection*{More on Rotation Operators}

The general unitary transformations is of the form:

\begin{equation}
\frac{1}{\sqrt {||a||^2 + ||b||^2}} \left[ \begin{array}{cc}
                                  a^* & b^* \\
                                  b & -a \\
                               \end{array} \right] 
\end{equation}
where $a^*$ and $b^*$ contain unknown phase angles.}

\end{document}